\documentclass[12pt]{article}

\usepackage[english]{babel}
\usepackage[utf8x]{inputenc}
\usepackage{amsmath, amsfonts, amssymb, amsthm}
\usepackage{graphicx}
\usepackage{cite}

\newtheorem{Thm}{Theorem}

\title{Heterogeneous epidemic model for assessing data dissemination in opportunistic networks}
\author{Vadim Alexeev, Liudmila Rozanova, Alexander Temerev}

\begin{document}
\maketitle

\begin{abstract}

In this paper we investigate a susceptible-infected-susceptible (SIS) epidemic model describing data dissemination in opportunistic networks with heterogeneous setting of transmission parameters. We obtained the estimation of the final epidemic size assuming that amount of data transferred between network nodes possesses a Pareto distribution, implying scale-free properties. In this context, more heterogeneity in susceptibility means the less severe epidemic progression, and, on the contrary, more heterogeneity in infectivity leads to more severe epidemics --- assuming that the other parameter (either heterogeneity or susceptibility) stays fixed. The results are general enough and can be useful in a broader context of epidemic theory, e.g. for estimating the progression for diseases with no significant acquired immunity --- in the cases where Pareto distribution holds.

\end{abstract}

\section{Introduction}

Investigating epidemic spreading in heterogeneous populations is a fairly hot topic today, as obtaining analytical results in such models can be somewhat challenging. In 2012, Artem S. Novozhilov provided an analytical proof ~\cite{novozhilov2012epidemiological} that for SIR (Susceptible-Infected-Recovered model) heterogeneity of susceptibility is inversely proportional to epidemic spreading speed, and heterogeneity of infectivity is proportional to spreading speed, and derived a formula for estimating the final epidemic size. 

In 2013, Benjamin Morin had suggested that extending these results for SIS (Susceptible-Infected-Susceptible model) is not feasible, as ``the reentry into the susceptible class causes the distributed equations to be completely unsolvable in any meaningful way''~\cite{morin2013variable}. However, we were able to achieve this (for a specific, though common, distribution), provided that either susceptibility or infectivity is fixed, and the other parameter remains heterogeneous.

\section{SIS model and its relevance}

SIS model describes the progress of an epidemic where there is no long lasting immunity. This can be applied to some infections such as those responsible for common cold, or data dissemination processes in opportunistic networks ~\cite{conti2013semantic}, or the spreading of information through gossip \cite{eugster2004epidemic}. For the simplest case when we assume that the population is closed, the contacting individuals are moving homogeneously, and infectivity and susceptibility are the same and constant for all individuals, the epidemic process is defined by the standard SIS model:

\begin{align}
\frac{dS}{dt}&=-\beta SI+\gamma I, \nonumber \\
\frac{dI}{dt}&=\beta SI-\gamma I. \nonumber
\end{align}

Denoting with $N$ the total number of contacting individuals, it holds that 

\begin{align}
\frac{dS}{dt}+\frac{dI}{dt}&=0,\Rightarrow S(t)+I(t)=N, \nonumber
\end{align}

it follows that 

\begin{align}
\frac{dI}{dt}=(\beta N-\gamma)I-\beta I^2,\nonumber
\end{align}

i.e. the dynamics of infection carriers is described by logistic equation, so for any start conditions where $I(0)>0$:

\begin{align}
\frac{\beta N}{\gamma}\le 1 \Rightarrow \lim_{t \to +\infty} I(t)=0. \nonumber
\end{align}

However, for most real-world processes the assumption of constant infectivity and constant susceptibility is too simplistic. For opportunistic networks, the trasmission rates of network nodes can be variable (which leads to distributed infectivity). In modeling the spreading of an infectious disease, the individual resistance to the disease can vary (hence distributed susceptibility). Occasionally, both parameters can be heterogeneous.

Therefore we can consider infectivity and susceptibility parameters to be functions, and assume that these parameters have distribution of the parameter in population; this method seems to be more promising.

Indeed, considering different algorithms of data transmission in opportunistic networks we see that the rate of this process is determined individually for each user, depending on the many factors. So, we can assume that users’ rates of data transfer have some sets of values with continuous distributions.

\section{Constructing data dissemination model in the terms of epidemic theory for heterogeneous population}

Let’s give a standard description of the data transfer process in terms of the epidemic theory for heterogeneous populations.

Let’s $s(t,\omega_{1})$ and $i(t,\omega_{2})$ be the densities of the susceptible (e.g. willing to accept data) and infectious (transmitting data) nodes respectively. The number of susceptible and infected nodes is described by the following function: 

\begin{align}
S(t)&=\int_{\Omega_{1}}{s(t,\omega_{1})}d\omega_{1},\nonumber \\
I(t)&=\int_{\Omega_{2}}{i(t,\omega_{2})}d\omega_{2}. \nonumber
\end{align}

where $\Omega_{1}$ and $\Omega_{2}$ are the sets of values of parameter $\omega_{1}, \omega_{2}$.

For these two subsets the probability density functions (PDFs) are given by

\begin{align}
p_{1}(t,\omega_{1})=\frac{s(t,\omega_{1})}{S(t)}, p_{2}(t,\omega_{2})=\frac{i(t,\omega_{2})}{I(t)}. \nonumber
\end{align}

In general form the dynamics of such system is described by the following expressions:

\begin{align}
\label{eq:mm1}
\frac{\partial}{\partial t}s(t,\omega_{1})&=s(t,\omega_{1})\cdot F_{1}(s(t,\omega_{1}),i(t,\omega_{2})), \nonumber \\
\frac{\partial}{\partial t}i(t,\omega_{2})&=i(t,\omega_{2})\cdot F_{2}(s(t,\omega_{1}),i(t,\omega_{2})).
\end{align}

the initial conditions are

\begin{align}
\label{eq:mm2}
s(0,\omega_{1})=S_{0}p_{1}(0,\omega_{1}), i(0,\omega_{2})=I_{0}p_{2}(0,\omega_{2}) . \nonumber
\end{align}

The functions $F_{1}$ and $F_{2}$ define the dynamics of interacting subgroups of susceptible and infectious users, which depends on the time, parameters, densities of both user’s subsets and their sizes. The important fact is that these functions cannot depend on the density functions explicitly, and have a special form. Applying the theory of heterogeneous populations\cite{novozhilov2012epidemiological}, we assume that this form must satisfy some additional requirements. In particular, we assume that:

\begin{align}
F_{i}(s(t,\omega_{1}),i(t,\omega_{2}))=f_{i}(S,I,\overline{\varphi}_{1}(t),\overline{\varphi}_{2}(t))+\overline{\varphi}_{i}(\omega_{i})g_{i}(S,I,\overline{\varphi}_{1}(t),\overline{\varphi}_{2}(t)),
\end{align}

where $\varphi_{i},f_{i},g_{i}$ are given functions, 

\begin{align}
\overline{\varphi}_{i}(t)=\int_{\omega_{i}}{\varphi_{i}(\omega_{i})p_{i}(t,\omega_{i})}\mathrm{d}\omega_{i}, i=1,2.\nonumber
\end{align}

Integrating the first equation in \eqref{eq:mm1} for $\omega_{1}$ and the second one for $\omega_{2}$, and using \eqref{eq:mm2}, we obtain the following system:

\begin{align}
\label{eq:mm3}
\dot{S}&=S\cdot[f_{1}(S,I,\overline{\varphi}_{1}(t),\overline{\varphi}_{2}(t))+\varphi_{1}(t)g_{1}(S,I,\overline{\varphi}_{1}(t),\overline{\varphi}_{2}(t))], \nonumber \\
\dot{I}&=I\cdot[f_{2}(S,I,\overline{\varphi}_{1}(t),\overline{\varphi}_{2}(t))+\varphi_{2}(t)g_{2}(S,I,\overline{\varphi}_{1}(t),\overline{\varphi}_{2}(t))].
\end{align}

Thus, the dynamics of this system depends only on the total size of $I$ and $S$ and $\overline{\varphi}_{i}(t)$ and can be obtained explicitly if we know $\overline{\varphi}_{i}(t)$.

Let's introduce auxiliary variables $q_{i}(t)$ as the solutions of the differential equations 

\begin{align}
\frac{dq_{i}(t)}{dt}=g_{i}(S,I,\overline{\varphi}_{1}(t),\overline{\varphi}_{2}(t)), i=1,2. \nonumber
\end{align}

From the results of Karev and others \cite{karev2005dynamics}, the current means of $\varphi_{i}(\omega_{i})$ are determined by the formulae:

\begin{align}
\overline{\varphi}_{i}(t)=\left.\frac{dM_{i}(0,\lambda)}{d\lambda}\right |^{}_{\lambda=q_{i}(t)}\frac{1}{M_{i}(0,q_{i}(t))}\nonumber
\end{align}

and satisfy the conditions:

\begin{align}
\frac{d}{dt}\overline{\varphi}_{i}(t)=g_{i}(S,I,\overline{\varphi}_{1}(t),\overline{\varphi}_{2}(t))\sigma_{i}^2(t),\nonumber
\end{align}

where $\sigma_{i}^2(t)$, are the current variances, $M_{i}(0,\lambda)$ are the moment generating functions (MGFs) of the initial distributions of $\overline{\varphi}_{i}(t,\omega_{i})$, $i=1,2$.

So, the analysis of infinite-dimensional model \eqref{eq:mm1} is reduced to the analysis of ODE system \eqref{eq:mm3} with small number of dimensions, and we need to know just the MGFs of the initial distributions.

Analytical investigation of this system gives very interesting results. In particular, composition of the population changes over time such a way that individuals with a lower value of the parameter functions are replaced by individuals with a high value of the parameter function.

\subsection{SIS model}

Suppose now that susceptibility and infectivity are distributed heterogeneously.
Let’s $\beta_{1}(\omega_{1})$ is the transmission parameter that encompasses the information on the probability of a successful contact and the contact rate for susceptible users, $\beta_{2}(\omega_{2})$ is the transmission parameter for infected users. Simplifying, we assume that the parameters of the two those subgroups are independent, i.e., 

\begin{align}
\beta(\omega_{1},\omega_{2})=\beta_{1}(\omega_{1})\beta_{2}(\omega_{2}).\nonumber
\end{align}

Also, let’s note that nothing else except for the standard law of mass action is supposed to define the rates of change in susceptible and infectious subpopulations, population is closed (i.e. the total number of contacting individuals is constant and equals to $N$) and the duration of ``informational infection'' (i.e. time of transmission) is distributed exponentially with mean $\frac{1}{\gamma}$.

The number of non-infected users with the value of susceptibility $\omega_{1}$, which are infected by users with the infectivity value $\omega_{2}$, is given by 

\begin{align}
\beta_{1}(\omega_{1})s(t,\omega_{1})\beta_{2}(\omega_{2})i(t,\omega_{2}).\nonumber
\end{align}

The total change in the infectious subpopulation with parameter value $\omega_{2}$ is

\begin{align}
\beta_{2}(\omega_{2})i(t,\omega_{2})\int_{\Omega_{1}} \beta_{1}(\omega_{1})s(t,\omega_{1})d\omega_{1}.\nonumber
\end{align}

Analogical expression can be obtained for infected sub-population:

\begin{align}
\beta_{1}(\omega_{1})s(t,\omega_{1})\int_{\Omega_{2}} \beta_{2}(\omega_{2})i(t,\omega_{2})d\omega_{2}.\nonumber
\end{align}

Combining the above assumptions we obtain the following SIS model for heterogeneous population:

\begin{align}
\frac{\partial}{\partial t}s(t,\omega_{1})=-\beta_{1}(\omega_{1})s(t,\omega_{1})\int_{\Omega_{2}} \beta_{2}(\omega_{2})i(t,\omega_{2})d\omega_{2}+\gamma i(t,\omega_{2}) \nonumber \\
=-\beta_{1}(\omega_{1})s(t,\omega_{1})\overline{\beta}_{2}(t)I(t)+\gamma i(t,\omega_{2}),\nonumber \\
\frac{\partial}{\partial t}i(t,\omega_{2})=-\beta_{2}(\omega_{2})i(t,\omega_{2})\int_{\Omega_{1}} \beta_{1}(\omega_{1})s(t,\omega_{1})d\omega_{1}-\gamma i(t,\omega_{1}) \nonumber \\
=\beta_{2}(\omega_{2})i(t,\omega_{2})\overline{\beta}_{1}(t)S(t)-\gamma  i(t,\omega_{1}).
\end{align}

where

\begin{align}
\overline{\beta}_{1}(t)=\int_{\Omega_{1}} \beta_{1}(\omega_{1})p_{s}(t,\omega_{1})d\omega_{1},\nonumber \\
\overline{\beta}_{2}(t)=\int_{\Omega_{2}} \beta_{2}(\omega_{2})p_{i}(t,\omega_{2})d\omega_{2}.
\end{align}

The model should be supplemented with initial conditions

\begin{align}
s(0,\omega_{1})=S_{0}p_{s}(0,\omega_{1}),
i(0,\omega_{1})=I_{0}p_{i}(0,\omega_{2}).\nonumber
\end{align}

where $p_{s}(0,\omega_{1}), p_{i}(0,\omega_{2})$ are the PDFs in the initial time $t=0$.

\paragraph{} According to the mechanism of reduction mentioned above, we can change this infinite-dimensional system to the equivalent ordinary system of differential equations:

\begin{align}
\label{eq:odes}
\dot{S}&=-\overline{\beta}_{1}(t)\overline{\beta}_{2}(t)SI+\gamma I, \nonumber \\
\dot{I}&=\overline{\beta}_{1}(t)\overline{\beta}_{2}(t)SI-\gamma I, \nonumber \\
\dot{q}&_{1}(t)=-\overline{\beta}_{2}(t)I, \nonumber \\
\dot{q}&_{2}(t)=\overline{\beta}_{1}(t)S.
\end{align}

where

\begin{align}
\label{mm7}
\overline{\beta}_{i}(t)=\partial_{\lambda}\ln M_{i}(0,\lambda) |_{\lambda=q_{i}(t)}.
\end{align}

$M_{i}(0,\lambda)$ are the MGFs in the time $t=0$, $i=1,2$.

\paragraph{Theorem 1.} Let $S^{(1)}(t),S^{(2)}(t)$ be the solutions of \eqref{eq:odes} with the initial conditions that satisfy $(\sigma^{(1)}_1)^2(0)>(\sigma^{(2)}_1)^2(0)$ for the distributions of susceptibility, all other initial conditions being equal. Then there exists an $\varepsilon>0$ such that $S_{1}(t)>S_{2}(t)$ for all $t\in(0,\varepsilon)$.

\paragraph{Proof.} Differentiating the first equation in the system \eqref{eq:odes} with fixed infectivity, we obtain

\begin{align}
\ddot S = -\dot{\overline{{\beta}}}_{1}(t)\overline{\beta}_{2}(t)SI - \overline{\beta}_{1}(t)\dot{\overline{\beta}}_{2}(t)SI - \overline{\beta}_{1}(t)\overline{\beta}_{2}(t)\dot{S}I - \overline{\beta}_{1}(t)\overline{\beta}_{2}(t)S \dot{I} +\gamma \dot{I}.
\end{align}

At $t=0$ we observe that $\dot{\overline{{\beta}}}_{1}(0)$ is proportional to $\sigma^2_1(0)$ and all other summands are equal by assumption. Thus, at the initial time moment, $\ddot{S}_{1}(0)>\ddot{S}_{2}(0)$ which proves the first part. The second part is proved in a similar way.

\paragraph{} 
This means that the more heterogeneous the susceptible hosts the less severe the disease progression (under the condition of fixed infectivity) and the opposite proposition (if susceptibility is fixed) is true: the more heterogeneous the infective class in infectivity, the more severe the disease progression.

\subsection{Exact solution for SIS model}
The system \eqref{eq:odes} could be represented as the nonlinear second order differential equation with variable coefficients:

\begin{align}
\label{eq:mm9}
\dot{I}&=\beta(t) I\left(I-\frac{\beta(t)N-\gamma}{\beta(t)}\right),  
\end{align}

where $\beta(t)=-\overline{\beta}_{1}(t)\overline{\beta}_{2}(t)$ for which \eqref{mm7} holds.

Let's $\beta(t) N -\gamma=\alpha (t)$, then (9) could he rewritten as:

\begin{align}
\dot{I}&+\alpha (t)I=\beta(t)I^2,  
\end{align}

and we can see that this is the Bernoulli differential equation with known exact solutions.

Indeed, dividing all terms of the equation by $I^2$, we obtain

\begin{align}
\label{eq:mm11}
\dot{I}I^{-2}+\alpha (t)I^{-1}=\beta(t), 
\end{align}

Performing the substitution $z(t)=I^{-1}$ and differentiating, we have

\begin{align}
\frac{dz}{dt}=-I^{-2}\frac{dI}{dt}.  \nonumber
\end{align}

Then, the equation \eqref{eq:mm11} can be reduced to the linear form:

\begin{align}
\frac{dz}{dt}-\alpha(t)z=-\beta(t), \nonumber
\end{align}

or, going back to the original notation,

\begin{align}
\label{eq:mm12}
\frac{dz}{dt}=(\gamma-\overline{\beta}_{1}(t)\overline{\beta}_{2}(t)N)z+\overline{\beta}_{1}(t)\overline{\beta}_{2}(t).  
\end{align}

The differential equation \eqref{eq:mm12} can be solved by Lagrange (continuous variation) method or by the method of integrating factors.
Thus, to find the exact solutions of the system \eqref{eq:odes}, we only need to know the value of the functions $\overline{\beta}_{1}(t)$ and $\overline{\beta}_{2}(t)$, that is the MGF of parameters susceptibility and infectivity. 

In general form the solution of this equation can be found as:

\begin{align}
I(t)=\frac{\exp{\int_1^t(\gamma-\beta(\phi)N)d\phi}}{C-\int_1^t\beta(\epsilon)\exp{\int_1^\epsilon(\gamma-\beta(\phi)N)d\phi}d\epsilon}.  
\end{align}

In the case with the constant transmission coefficients all non-trivial solutions of the equation \eqref{eq:mm9} tend to the equilibrium $I(t)=\frac{\beta N-\gamma}{\beta}$ if $t \rightarrow +\infty$. But for the variable coefficients case this doesn't hold. However, if the improper integrals $\int_1^t(\gamma-\beta(\phi)N)d\phi$ and $\int_1^t\beta(\epsilon)\exp{\int_1^\epsilon(\gamma-\beta(\phi)N)d\phi}d\epsilon$ diverge when $t \rightarrow +\infty$, then all non-zero solutions tend to the equilibrium.

\subsection{On the final epidemic size with one distributed transmission parameter}

Theorem 2.1 in \cite{novozhilov2012epidemiological} makes some assumptions on the existence of the MGFs for the initial distribution. 

In particular, $M_{i}(0,\lambda)$ are the MGFs in the time $t=0$, $i=1,2$, as long as the derivatives on the right hand side of \eqref{mm7} exist. Dealing with distributions of more general nature, where MGFs exist only for some values of the parameter, requires more care.

It's clear that $\overline{\beta}_{i}(t) \ge 0$, $I \ge 0$, $S\ge 0$. Thus, according to the last two equations of the above system, the quantities $q_1(t)$ and $q_2(t)$ are monotonically decreasing resp. increasing. The initial values vanish: $q_1(0)=q_2(0) = 0$; thus $q_1(t) \le 0$, $q_2(t)\ge 0$ for all $t\ge 0$.

This allows us to prove the following:
\begin{Thm}
 Let $\beta_2$ is a constant and the initial distribution of $\beta_1$ be a distribution with the moment generating function $M(\lambda)$, $\lambda\le 0$, such that $H(\lambda) = \partial_\lambda \ln M(\lambda)$ has finite limits for $\lambda\nearrow 0$ and $\lambda \to -\infty$.
 
 Suppose that there is a limit
\[
 I_\infty = \lim_{t\to +\infty} I(t) \neq 0.
\]

Then the following equation holds:
\[
 \beta_2 \chi S_\infty = \gamma,
\]
where $\chi = \lim\limits_{\lambda\to\-\infty} H(\lambda)$.
 \end{Thm}
An example of a probability distribution with these properties is Pareto distribution with starting value $\xi$ and degree $\alpha > 1$. The MGF of the Pareto distribution exists only for $\lambda \le 0$ and is equal to
\[
 M(\lambda) = \alpha(-\xi\lambda)^\alpha\Gamma(-\alpha,-\xi\lambda),\quad \lambda < 0,
\]
extended by continuity to $\lambda = 0$.

We obtain
\[
 H(\lambda)=\partial_\lambda \ln M(\lambda) = -\frac{\alpha}{\lambda} + \frac{\xi e^{\xi\lambda}(-\xi\lambda)^{-\alpha-1}}{\Gamma(-\alpha,-\xi\lambda)}, 
\]
which has a finite limit
\[
 \frac{\xi\alpha}{\alpha-1}
\]
as $\lambda \nearrow 0$ and a finite limit  $\xi$ as $\lambda\to-\infty$.

\begin{proof}
The system of ODEs \eqref{eq:odes} can be derived by repeating the proof of Theorem 1 in \cite{novozhilov2012epidemiological}, using the fact that $q_1(t)\le 0$ for all $t\ge 0$ and extending $H(\lambda)$ by continuity to $\lambda = 0$ if needed.

Just as in Theorem 4 of \cite{novozhilov2012epidemiological}, we use the first integrals of the system to obtain the final epidemic size. Indeed, from the second equation we obtain
\[
 \frac{d}{dt}\ln I = \beta_2 \dot q_2 - \gamma,
\]
thus
\[
 \frac{I}{I_0} = e^{\beta_2 q_2(t)-\gamma t}.
\]

From this we immediately get
\[
 \dot q_1 = -\beta_2 I_0 e^{\beta_2 q_2(t)-\gamma t},
\]
\[
 \dot q_2 = H(q_1(t))(N-I_0e^{\beta_2 q_2(t)-\gamma t}).
\]

Suppose now there is a limit
\[
 I_\infty = \lim_{t\to +\infty} I(t) \neq 0.
\]

Then, using that $H(\lambda)\to \chi$, $\lambda\to-\infty$, we get
\[
\dot q_2(t) \to N-I_\infty = S_\infty \chi. 
\]
On the other hand, $\beta_2 q_2(t) - \gamma t$ and its derivative both have finite limits, and therefore the limit of the derivative is equal to 0:
\[
 \beta_2 \chi S_\infty = \gamma.
\]
\end{proof}

\section {Results}

We constructed and analytically investigatef a specific epidemic model: SIS model with heterogeneous susceptibility and infectivity, giving particular attention to the case with Pareto distribution of these parameters, which holds a practical importance for most real world cases possessing scale-free properties \cite{albert2002statistical}.

Such type of models may be used for describing data dissemination in networks with heterogeneous transmission parameters, for example, for the opportunistic networks with semantic routing algorithms proposed in \cite{conti2013semantic}.

Using specific assumptions about the MGF function form of transmission parameters (based on special properties of scale-free networks), we have analyzed the model behavior and obtained the formula for estimating the final epidemic size.

We have shown that in this context, more heterogeneity in susceptibility means the less severe epidemic progression, and, on the contrary, more heterogeneity in infectivity leads to more severe epidemics.

These results are general enough to be used in other fields where SIS model is applicable (and Pareto distribution of susceptibility / infectivity holds), like estimating the epidemic progression for diseases with no significant acquired immunity, like common cold.

\bibliographystyle{unsrt}
\bibliography{main}{}

\end{document}